\documentclass[twocolumn,preprint,preprintnumbers,showpacs,floatfix]{revtex4}
\usepackage{times}
\usepackage{epsfig}
\usepackage{amssymb}
\usepackage[english]{babel}
\usepackage{graphics}

\begin{document}
\bibliographystyle{apsrev}


\title{A generalized approach to complex networks}
\author{L. da F. Costa and Luis Enrique C. da Rocha}
\affiliation{ Institute of Physics of S\~ao Carlos, University of S\~ao Paulo \\
              Caixa Postal: 369, CEP: 13560-970, S\~ao Carlos SP, Brazil. \\
              luciano@if.sc.usp.br and rocha@if.sc.usp.br}
\date{Received: 28th September 2005}

\begin{abstract} This work describes how the
formalization of complex network concepts in terms of discrete
mathematics, especially mathematical morphology, allows a series of
generalizations and important results ranging from new measurements of
the network topology to new network growth models.  First, the
concepts of node degree and clustering coefficient are extended in
order to characterize not only specific nodes, but any generic
subnetwork. Second, the consideration of distance transform and rings
are used to further extend those concepts in order to obtain a
signature, instead of a single scalar measurement, ranging from the
single node to whole graph scales.  The enhanced discriminative
potential of such extended measurements is illustrated with respect to
the identification of correspondence between nodes in two complex
networks, namely a protein-protein interaction network and a perturbed
version of it.  The use of other measurements derived from
mathematical morphology are also suggested as a means to characterize
complex networks connectivity in a more comprehensive fashion.

\end{abstract}

\maketitle

\section{Introduction}

\label{sec:1} One of the unavoidable consequences of the fast pace
of developments in the new area of complex networks
\cite{Albert_Barab:2002, Dorog_surv, Dorog_book, Newman03} is that,
while many impressive and relevant concepts and perspectives have been
identified and well-developed, and promising results been obtained,
some interesting issues have received relatively little attention.
One particularly important point is the fact that, despite the major
advances achieved by using powerful tools from theoretical physics
(e.g.  \cite{Albert_Barab:2002, Newman03}), relatively little
attention has been given to the treatment of complex networks in terms
of discrete mathematics and mathematical morphology, which are
themselves well-established investigation fields.  Developed mainly by
J. Serra and collaborators \cite{Serra:1984} , the area of
mathematical morphology is aimed, through strict mathematical
formalization, at representing and analyzing the geometrical and
topological features of discrete mathematical structures, especially
regular lattices such as those underlying digital images
\cite{CostaCesar:2001}.  Mathematical morphology is strongly founded
on the discrete operations of complement, dilation and erosion, which
can be composed in order to obtain a whole series of new operators
with specific properties.  At the same time, previous developments by
L. Vincent and H. Heijmans \cite{Vincent:1989, Heij_etal:1992} have
shown how the mathematical morphology framework can be extended to
graphs, allowing not only the precise mathematical representation and
manipulation of those general structures, but also the immediate
access to the wealthy of existing results from mathematical
morphology.

The present article reports on how the application of discrete
mathematics, especially mathematical morphology \cite{Serra:1984} and
distance-oriented concepts \cite{Vincent:1989,
Heij_etal:1992,CostaCesar:2001}, bears the potential not only for
formalization, but also to obtain a series of new concepts and
results.  In particular, by considering the dilations of subnetworks
of a network $\Gamma$ and extending the concepts of numbers of
neighbors~\cite{faloutsos,faloutsos_conf,ego_centered,Havlin_tomo} and
hierarchical node degree~\cite{Costa_hier:2003}, we show that the
traditional concepts of node degree and clustering coefficient
\cite{Albert_Barab:2002, Newman03} can be generalized in two important
ways.  First, the concept of subnetwork dilation paves the way to
generalize the degree and clustering coefficient to any subnetwork of
$\Gamma$, and not only their specific nodes as adopted in the complex
network literature.  Such a concept therefore allows us to speak of
the degree of subgraphs of special interest, such as cycles, sets of
hubs, or the maximum spanning tree of a given complex network.
Second, the consideration of a series of subsequent dilations,
together with the respectively induced distance transform and rings,
allow the further extension of the degree and clustering coefficient
so that a signature, instead of the single scalar traditional
measurements, is obtained which can provide information about the
network connectivity from the node to the whole graph scales.

The potential of such hierarchical extensions for discriminating the
connectivity around each node (or subgraph) can be readily appreciated
by considering the fact that several nodes in a complex network will
have the same degree and clustering coefficient, but very few nodes
will share such values calculated for a series of subsequent
neighborhoods.  Such an interesting feature of the generalized
measurements is illustrated in the present article with respect to
protein-protein interaction networks.

\section{Basic Concepts}
\label{sec:2} 
\begin{figure*}
 \begin{center}
   \includegraphics[scale=0.25,angle=-90]{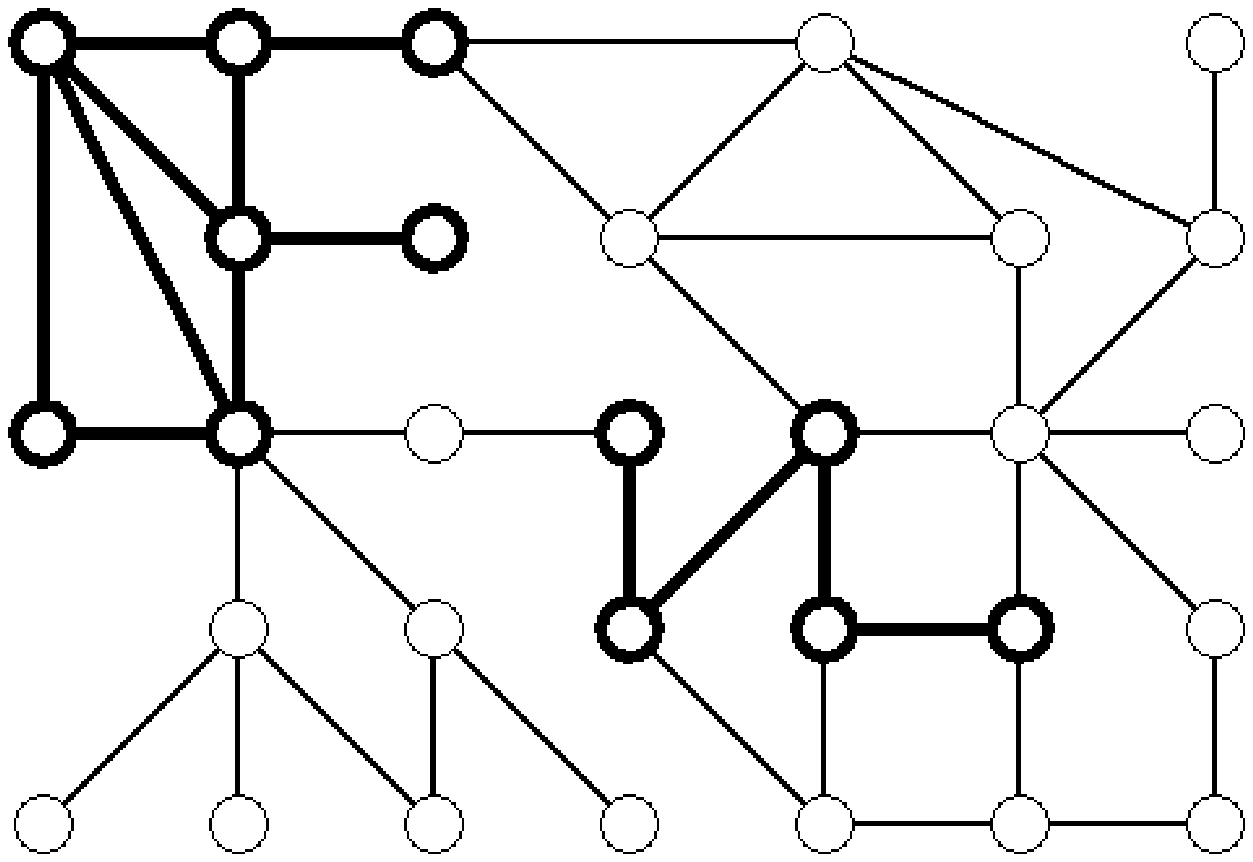}  \hspace{0.5cm}
   \includegraphics[scale=0.25,angle=-90]{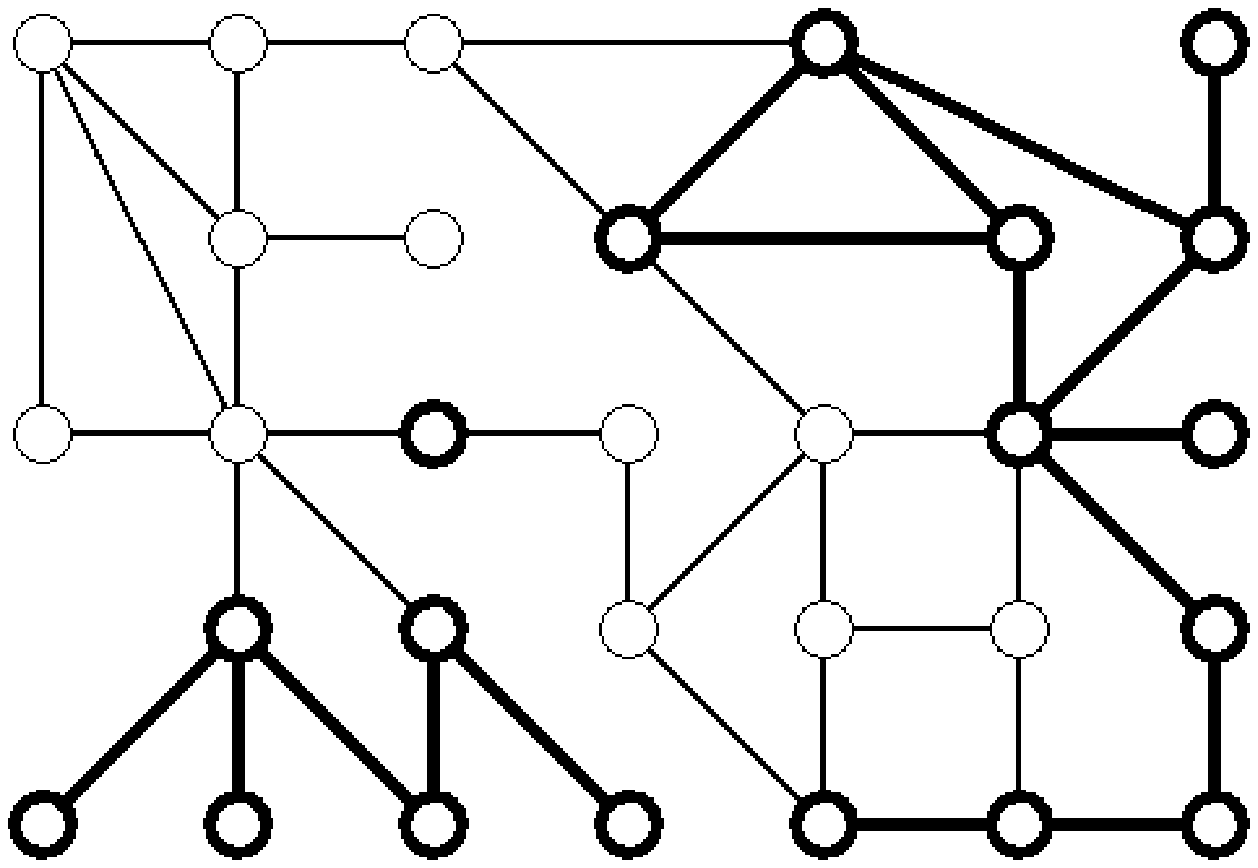}  \hspace{0.5cm}
   \includegraphics[scale=0.25,angle=-90]{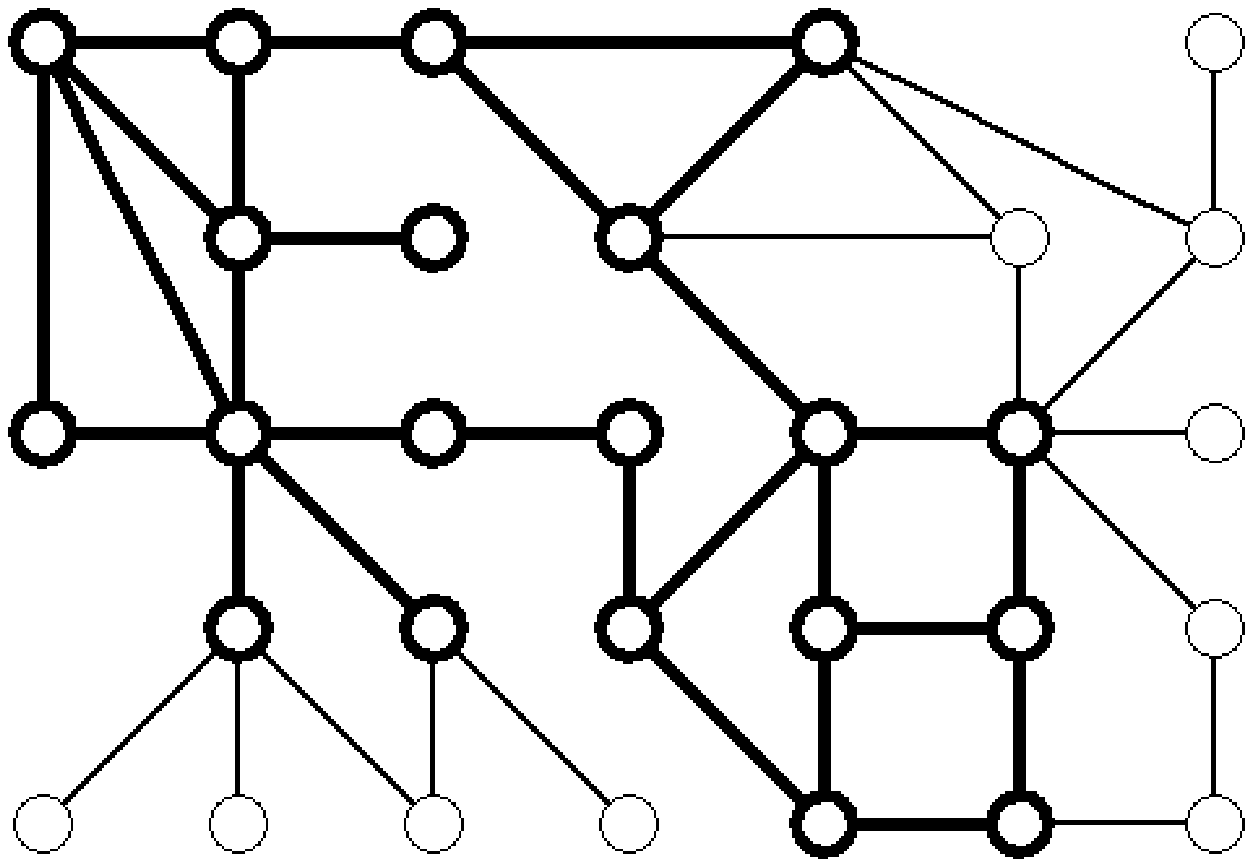}  \hspace{0.5cm}
   \includegraphics[scale=0.25,angle=-90]{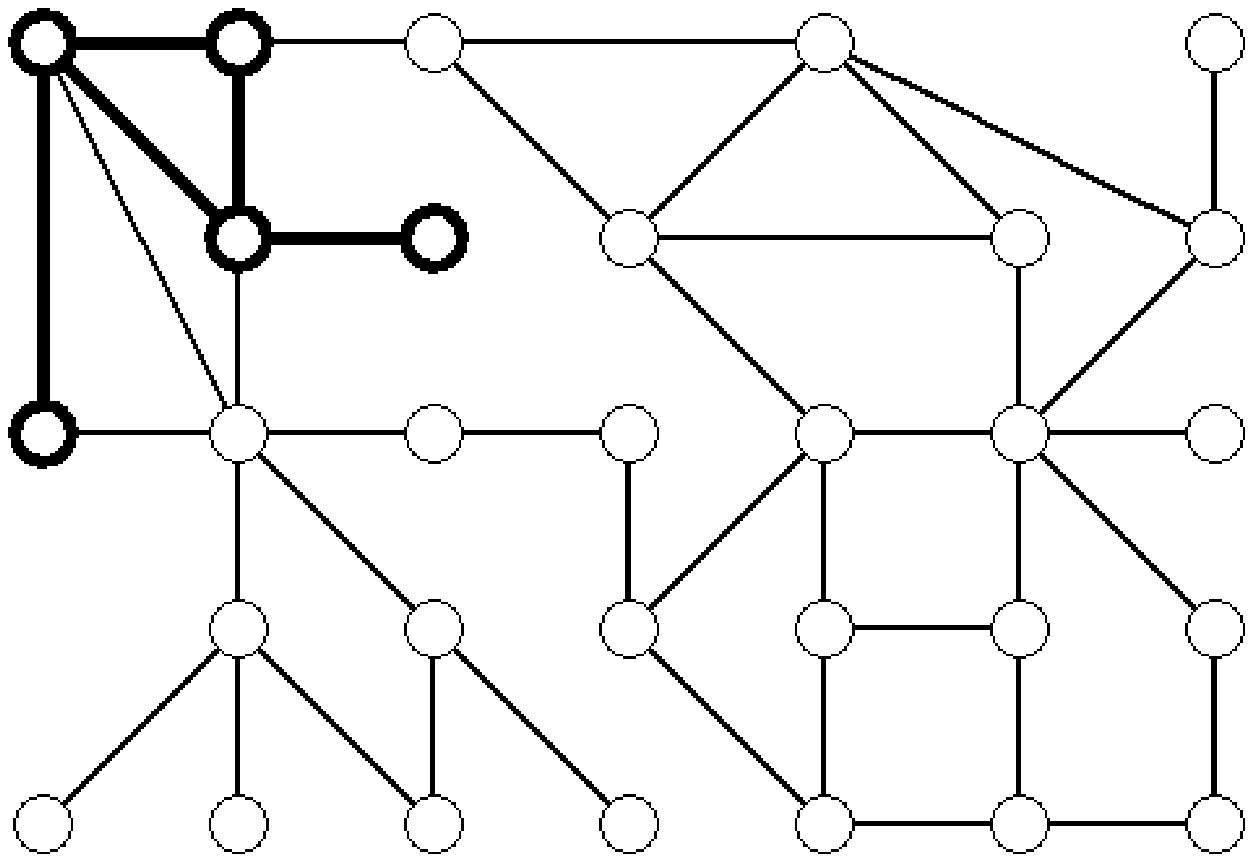}  \\
   (a) \hspace{3.5cm}   (b) \hspace{3.5cm} (c)  \hspace{3.5cm} (d) \\
   \includegraphics[scale=0.25,angle=-90]{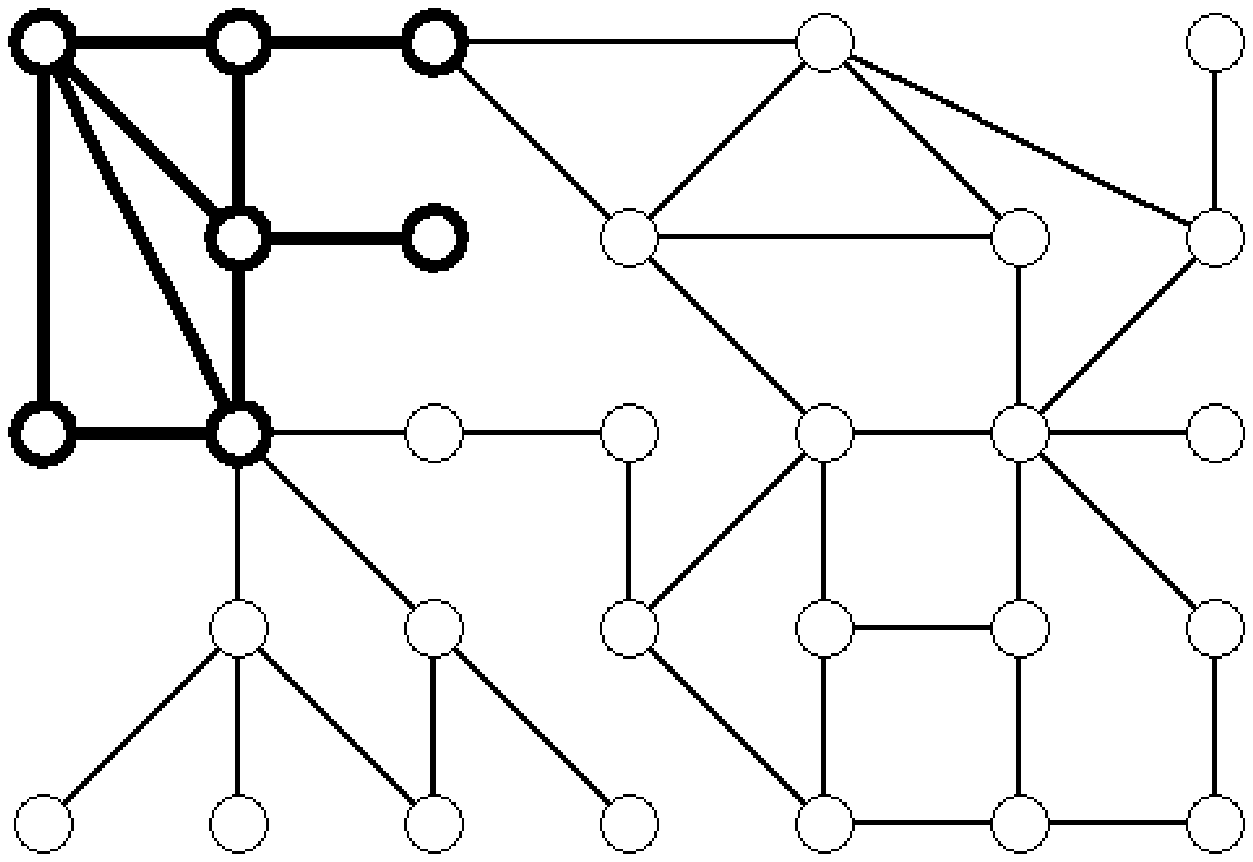}  \hspace{0.5cm}
   \includegraphics[scale=0.25,angle=-90]{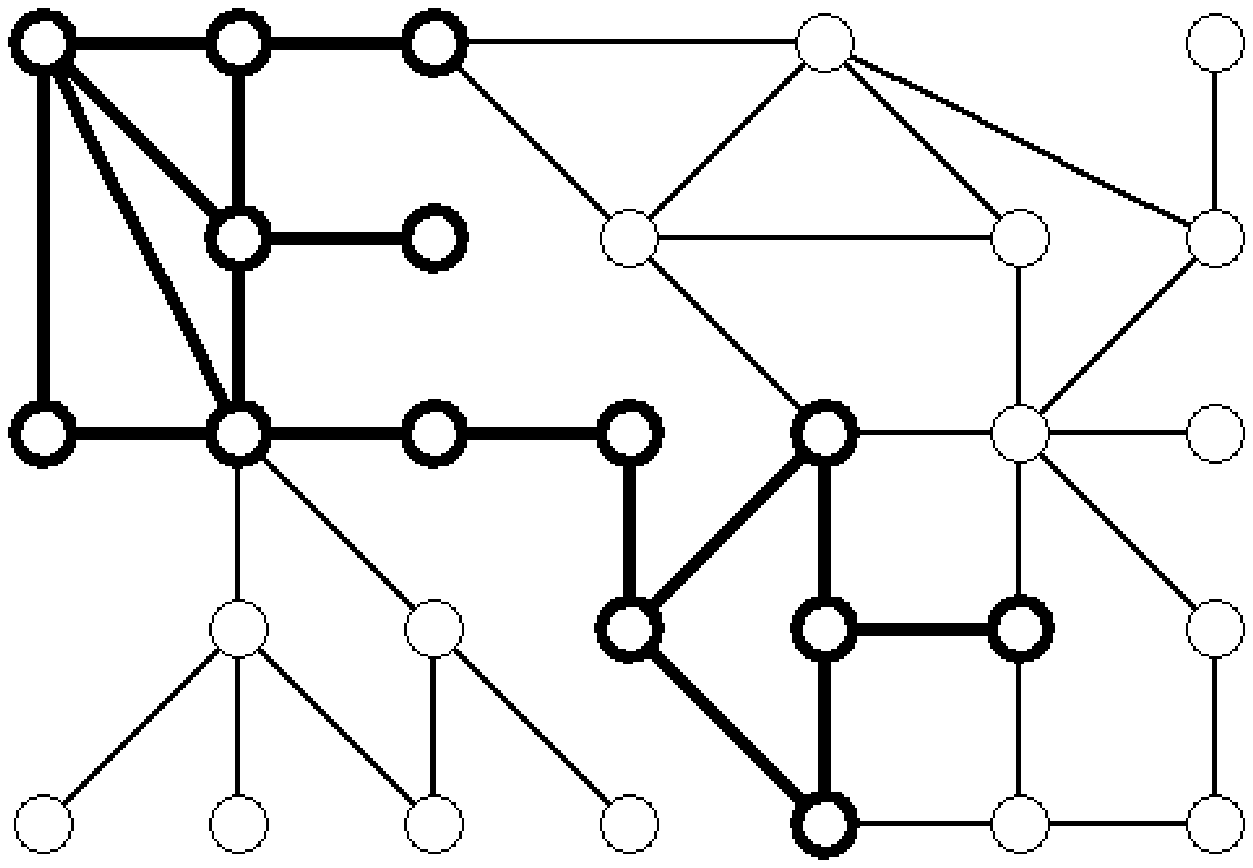}  \hspace{0.5cm}
   \includegraphics[scale=0.25,angle=-90]{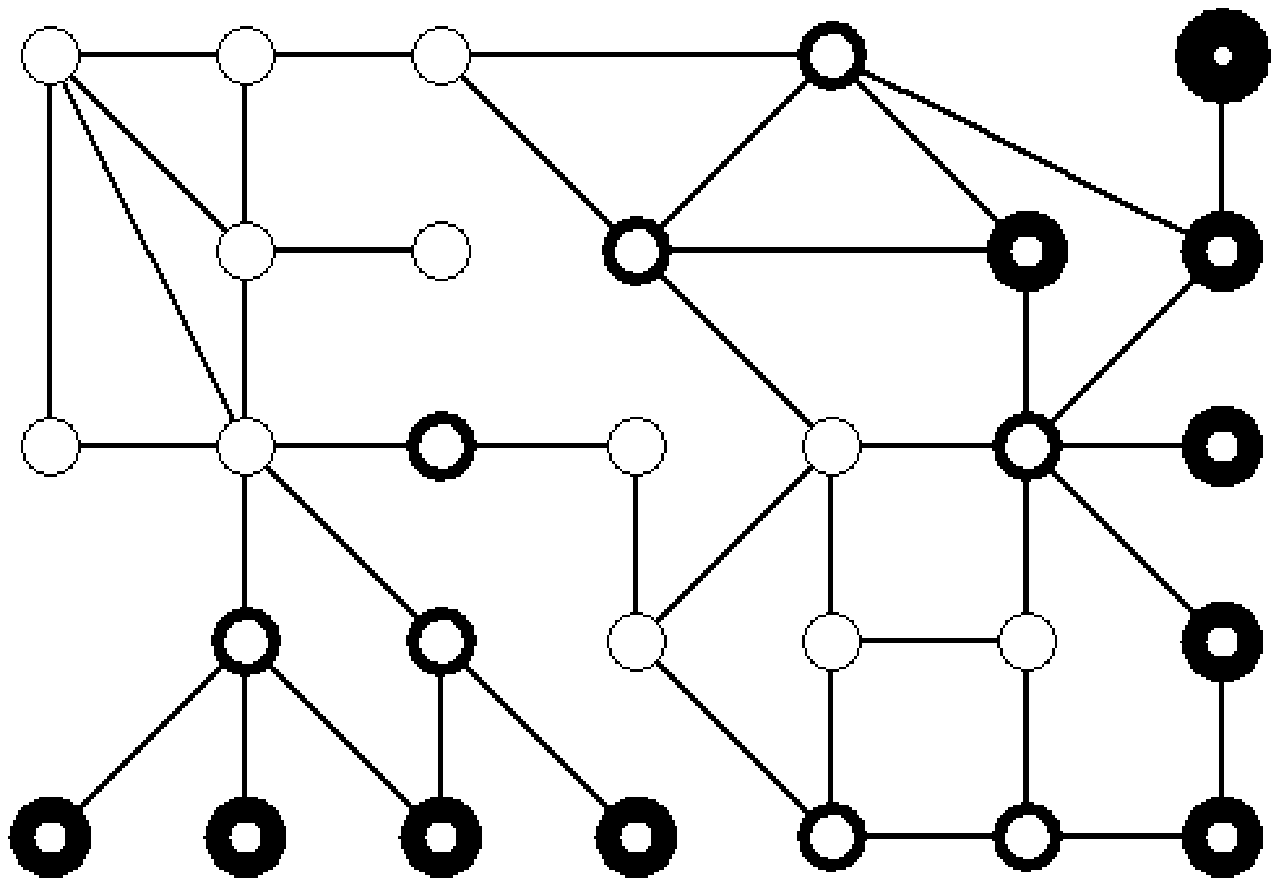}  \hspace{0.5cm}
   \includegraphics[scale=0.25,angle=-90]{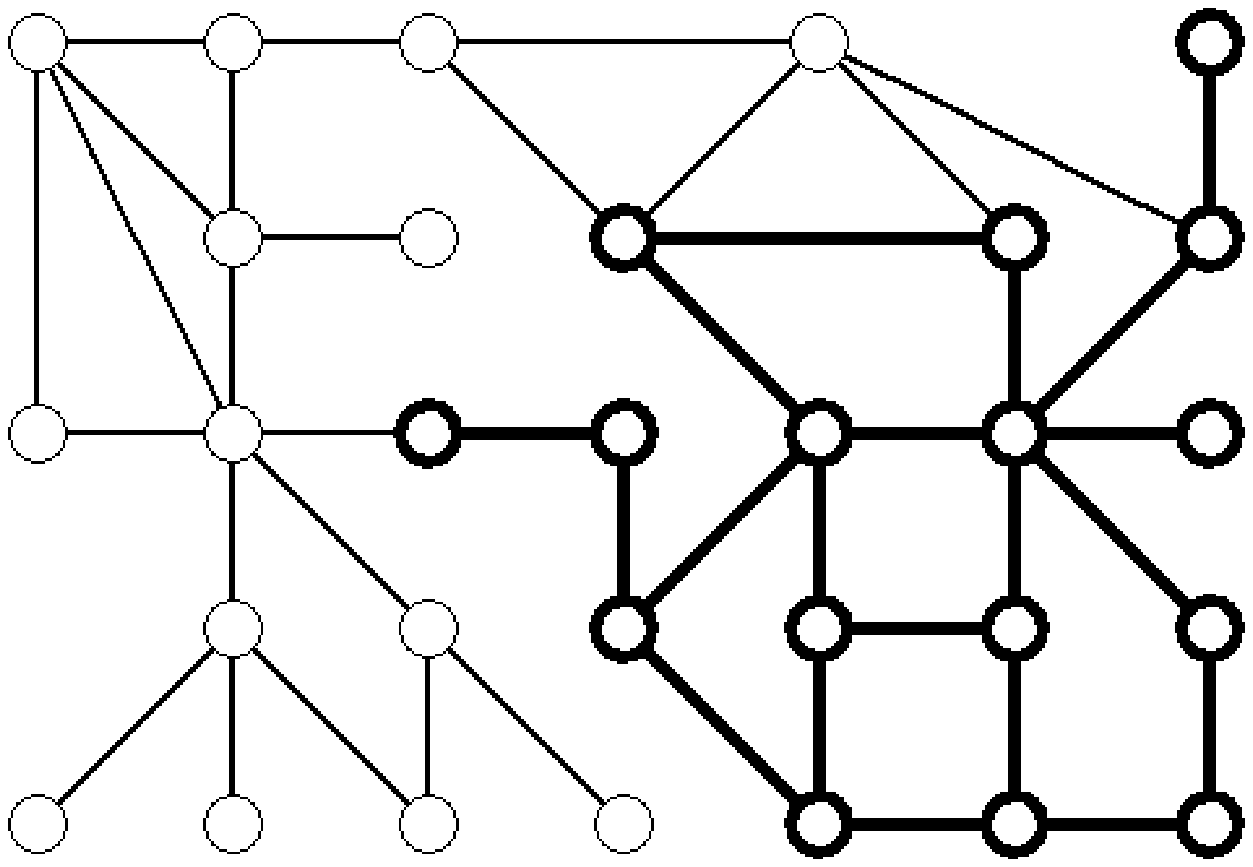}  \\
   (e) \hspace{3.5cm}   (f)  \hspace{3.5cm} (g) \hspace{3.5cm}   (h)

   \caption{Original network containing a subnetwork (a), identified
   by wider-border nodes, and its respective complement (b), dilation
   (c), erosion (d), opening (e), closing (f), and distance transform
   (g), with the distances identified by the node border width. The
   generalized Voronoi tessellation of the two connected components of
   the subnetwork in (a) is shown in (h).~\label{fig:ex1}}
\end{center}
\end{figure*}

A \emph{network} $\Gamma$ without multiple edges is a
discrete structure composed of a set of \emph{nodes} $V(\Gamma)$ and a
set $E(\Gamma)$ of \emph{edges} $(u, v)$ established between specific
pairs of nodes of $V(\Gamma)$, so that the network $\Gamma$ is
represented as $\Gamma=(V, E)$.  As we consider undirected networks
without loops, it follows that $(u, v) \iff (v, u)$ and $(u, u) \notin
E(\Gamma)$. Such a network can be conveniently represented in terms of
its respective \emph{adjacency matrix} $K$ such that each edge $(u,
v)$ is represented by making $K(u, v)=K(v, u)=1$, while the absence of
edge is indicated by zero value. A \emph{subnetwork} $\xi$ of $\Gamma$
is any network such that $V(\xi) \subseteq V(\Gamma)$ and $E(\xi)
\subseteq \left\{ (u, v) | (u, v) \in E(\Gamma) \: and \: u, v \in
V(\xi) \right\}$.  Figure~\ref{fig:ex1}(a) illustrates a network
$\Gamma$ and one of its many subnetworks $\xi$, identified by the
wider-border nodes and wider edges. Particularly interesting
subnetworks of a network $\Gamma$ include its hubs, outmost nodes
(i.e. nodes with low degree), as well as its cycles. Special cases of
subnetworks of $\Gamma$ include the empty network $(V=\emptyset,E=\emptyset)$,
where $\emptyset$ stands for the empty set, networks containing an isolated
node $u$ $\Gamma_u=(V=\left\{u \in V(\Gamma) \right\}, E=\emptyset)$, and
the own original network $\Gamma$.

The \emph{complement} of a subnetwork $\xi$ of $\Gamma$ is the
subnetwork $\xi'_\Gamma$ of $\Gamma$ such that $V(\xi'_\Gamma) =
\left\{ u | u \in V(\Gamma) \: and \: u \notin V(\xi) \right\}$ and
$E(\xi'_\Gamma) = \left\{ (u, v) | (u, v) \in E(\Gamma) \: and \: u, v
\in V(\xi'_\Gamma) \right\}$. Figure~\ref{fig:ex1}(b) illustrates the
complement $\xi'_\Gamma$ of $\xi$ in $\Gamma$.  A subnetwork is
\emph{connected} if any of its nodes can be reached from any of its
other nodes.  Two subnetworks $\zeta$ and $\xi$ of $\Gamma$ are
connected if it is possible to reach a node of $\xi$ from a node of
$\zeta$, and vice-versa. The maximal connected subnetworks, in the
sense of including the largest number of nodes, of a network are
called \emph{connected components}.  The subnetwork in
Figure~\ref{fig:ex1}(a) is not connected but contains two connected
components.

The \emph{degree} of a node $u$ of $\Gamma$, hence $k(u)$, corresponds
to the number of edges attached to that node.  The \emph{degree} of a
subnetwork $\xi$ of $\Gamma$, hence $k(\xi)$, is defined as the number
of edges implied by the dilation of $\xi$, i.e. those edges connecting
$\xi$ to the rest of $\Gamma$.  For instance, the degree of the
subnetwork $\xi$ in Figure~\ref{fig:ex1}(a) is 12.  The \emph{outmost
set} of a subnetwork $\xi$ of $\Gamma$ is the set of nodes
$\Omega(\xi)$ which have unit degree.  For simplicity's sake, such
nodes are henceforth referred to as \emph{outnodes}.  The
\emph{1-neighborhood} of a node $u$ of $\Gamma$, henceforth
represented as $n_1(u)$, is the set of nodes of $\Gamma$ which are
attached to $u$, plus node $u$.  This concept can be immediately
extended to express the \emph{neighborhood of a subnetwork} $\xi$ of
$\Gamma$, given as the set of nodes of $\Gamma$ which are connected to
$\xi$ plus the nodes in $V(\xi)$.

\section{Complex Network Morphology}
\label{sec:3}

The \emph{dilation} of a subnetwork $\xi$ of $\Gamma$ is defined as
the subnetwork $\delta(\xi)$ of $\Gamma$ having
$V(\delta(\xi))=n_1(\xi)$ as its set of nodes while its set of edges
include the edges of $\Gamma$ found between the nodes in
$n_1(\xi)$. The \emph{erosion} of $\xi$, represented as
$\varepsilon(\xi)$, is a subnetwork of $\Gamma$ which can be defined
as the complement of the dilation of $\xi'_\Gamma$,
i.e. $\varepsilon(\xi) = (\delta(\xi'_\Gamma))'_\Gamma$.  Observe that
the dilation or erosion of $\Gamma$ yields $\Gamma$ as result.
Figures~\ref{fig:ex1}(c) and (d) ilustrates the dilation and erosion
of the subnetwork $\xi$ in (a), respectively.  Observe that the
erosion eliminated one of the connected components of
$\xi$. Generally, $\delta(\varepsilon(\xi)) \neq
\varepsilon(\delta(\xi))$, i.e. the dilation can not be used to undo
an erosion, and vice-versa.  Observe also that a subnetwork $\xi$ is
necessarily contained or equal to its respective dilation, while the
erosion of a subnetwork $\xi$ is necessarily contained or equal to
itself.

The \emph{d-dilation} of a subnetwork $\xi$ is defined as the subnetwork
obtained by dilating $d$ times the subnetwork $\xi$, i.e.:

\begin{equation}
  \delta_d(\xi) = \underbrace{ \delta(\delta (\ldots (\xi) \ldots ))}_{d}
\end{equation}

Similarly, the \emph{d-erosion} can be defined as:

\begin{equation}
  \varepsilon_d(\xi) = \underbrace{ \varepsilon(\varepsilon (\ldots (\xi) \ldots ))}_{d}
\end{equation}

Observe that $\delta_d(\xi)$ converges to $\Gamma$ as $d$ is
increased, while $\varepsilon_d(\xi)$ converges to the empty network
under similar circumstances.  We also have that
$\delta_{d=i+j}(\xi)=\delta_i(\delta_j(\xi))=\delta_j(\delta_i(\xi))$
and
$\varepsilon_{d=i+j}(\xi)=\varepsilon_i(\varepsilon_j(\xi))=\varepsilon_j(\varepsilon_i(\xi))$.
The \emph{d-degree} of a subnetwork $\xi$ is defined as the degree of
the \emph{d}-dilation of the network $\xi$,
i.e., $k_d(\xi)=k(\delta_d(\xi))$.

It is possible to use combinations of dilations and erosions of a
subnetwork $\xi$ in order to obtain new operators such as the
\emph{opening} and \emph{closing} of $\xi$, which are defined as
$\alpha(\xi) = \delta(\varepsilon(\xi))$ and $\omega(\xi) =
\varepsilon(\delta(\xi))$, respectively.  Figures~\ref{fig:ex1} (e)
and (g) illustrate, respectively, the opening and closing of the
subnetwork $\xi$ in (a).  Observe that the closing of $\xi$ had as an
effect the connection of the two components of that subnetwork,
filling the gap between those subnetworks.  The opening and closing
operations are idempotent, in the sense that
$\alpha(\alpha(\xi))=\alpha(\xi)$ and
$\omega(\omega(\xi))=\omega(\xi)$.  It is also interesting to define
the \emph{d-opening} of a subnetwork $\xi$, henceforth represented as
$\alpha_d(\xi)$, corresponding to $d$ erosions followed by $d$
dilations.  The \emph{d-closing} of $\xi$, represented as
$\omega_d(\xi)$ can be defined in similar fashion.  The latter
operator is useful for investigating the progressive merging of
subnetworks of $\Gamma$ in terms of increasing values of $d$.
Particularly, interesting information about the network structure can
be provided by the evolution of the number of connected subnetworks,
starting from a specific set $\chi$ of subnetworks (e.g. the network
3-cycles), in terms of a sequence of \emph{d-}openings (or closings)
performed for increasing values of $d$.

\section{Distances, Distance Transforms, Parallels and Rings}
\label{sec:4}

Several important features of a complex network are related to the
concept of \emph{distance}.  If $\zeta$ and $\xi$ are any subnetworks
of $\Gamma$, the (minimal) distance between the respective set of
nodes $V(\zeta)$ and $V(\xi)$, hence $D(V(\zeta), V(\xi))$, can be
defined as the value of $d$ for which some node $u$ of $\zeta$ becomes
included into $V(\delta_d(\xi))$. It can be verified that $D(V(\zeta),
V(\xi)) = D(V(\xi), V(\zeta))$.  Observe that $D(V(\zeta),
V(\zeta))=0$.  In particular, the distance between a node $u$ and a
subnetwork $\xi$ is given as $D(\{ u \},V(\xi))$.  The \emph{distance
transform} of a subset of nodes $\chi$ of $\Gamma$ is the mapping which
assigns $D(\{ u \},\chi)$ to every node $u \in V(\Gamma)$, including
those in $\xi$. Figure~\ref{fig:ex1}(g) illustrate the distance
transform of the subnetwork in (a), with the distance values (4 values
for this subnetwork, i.e. $d = 0, 1, 2,$ and $3$) expressed in terms
of the node borders widths.  Given a subnetwork $\xi$ of $\Gamma$, the
subnetwork $\varrho$ defined by the set $V(p)$ of nodes such that
$D(V(\varrho),V(\xi))=d$ and the set of those edges of $\Gamma$
connecting nodes in $V(\varrho)$ is called the \emph{d-parallel} of
$\xi$, henceforth represented as $P_d(\xi)$.  The parallels of the
subnetwork $\xi$ in Figure~\ref{fig:ex1}(a) correspond to the set of
nodes with the same width in (g) plus the respective interconnecting
edges. The number of nodes and edges in a \emph{d-}parallel of $\xi$
are henceforth represented as $n \{ P_d(\xi) \}$ and $e \{ P_d(\xi)
\}$. Similarly, it is interesting to define the \emph{rs-ring} of
$\xi$, hence $R_{rs}(\xi)$, which corresponds to the union of the
respective parallels of $\xi$ for distances $d=r$ to $s$ plus the
edges of $\Gamma$ interconnecting such parallels.  The number of nodes
and edges in a \emph{rs}-ring of $g$ are henceforth represented as $n
\{ R_{rs}(\xi) \}$ and $e \{ R_{rs}(\xi) \}$.  Observe that a
\emph{d-}parallel therefore is the particular case of the
\emph{rs-}ring for $d=r=s$.  Another interesting possibility is to use
the above introduced distance concepts in order to obtain the
generalized Voronoi tessellation of subnetworks, as illustrated in
Figure~\ref{fig:ex1}(h) with respect to the two connected components
in the subnetwork in Figure~\ref{fig:ex1}(a).

The above definitions allow the concept of \emph{clustering
coefficient} \cite{Albert_Barab:2002, Newman03} to be
generalized to parallels and rings of any subnetwork.  The
\emph{rs-clustering coefficient} of a subnetwork $\xi$ of $\Gamma$,
henceforth represented as $cc_{rs}(\xi)$, can be defined as the
number of edges in the respective \emph{rs}-ring subnetwork,
divided by the total of possible edges between the nodes in that
ring, i.e.:

\begin{equation}
  cc_{rs}(\xi) = \frac{2e \{ R_{rs}(\xi) \} }{n \{ R_{rs}(\xi)
  \} (n \{ R_{rs}(\xi) \}-1)}
\end{equation}

\section{Hierarchical Measurements for Single Nodes}
\label{sec:5}

The concepts discussed above can be naturally extended to a single
node and to an edge, respectively, whether the subgraph contains the
node alone and whether the subgraph contains the edge and both nodes
connected by the edge \cite{Costa_Hier2:2005,Costa_Rodrigues:2005}.
Using the concept of rings considered in the last section, henceforth
the subgraph $\zeta$ composed of the ring $R_{d}(u)$ is defined as the
\emph{hierarchical level} related to the subgraph $\xi$ composed of
the single node $(u)$, such that, the \emph{hierarchical number of
nodes} $n_{d}(u)$ (or $n\{R_{d}(u) \}$) is given as the number of
nodes at \emph{hierarchical distance} $d$ from de reference node
$(u)$, i.e., the number of nodes in the ring $R_{d}(u)$. Hence, the
\emph{hierarchical degree} $k_{d}(u)$ is defined as the number of
edges between the nodes in the ring $R_{d}$ and $R_{d+1}$, such that
the \emph{hierarchical number of edges} among the nodes in the ring
$R_{d}(u)$ is $e_{d}(u)$ (or $e\{R_{d}(u)\}$). The \emph{hierarchical
clustering coefficient} $cc_{d}(u)$ is written using equation 3.
\begin{equation}
cc_{d}(u) = \frac{2e_{d}(u)}{n_{d}(u)(n_{d}(u)-1)}
\end{equation}
At last, other hierarchical measurements can be derived from the
above definitions and each of them will be dealt with in turn:
\begin{itemize}
\item \emph{Convergence ratio} $(C_{d}(u))$ : Measures the ratio between the hierarchical node degree of node $(u)$ at hierarchical distance $d-1$ and the hierarchical number of nodes in the ring $R_{d}(u)$.
\begin{equation}
C_{d}(u)=\frac{k_{d-1}(u)}{n_{d}(u)}
\end{equation}
\item \emph{Intra-ring degree} $(A_{d}(u))$ : The average among the degrees of the nodes in the ring $R_{d}(u)$.
\begin{equation}
A_{d}(u)= \frac{2e_{d}(u)}{n_{d}(u)}
\end{equation}
\item \emph{Inter-ring degree} $(E_{d}(u))$ : The average of the number of connections between each node in ring $R_{d}(u)$ and those in $R_{d+1}(u)$.
\begin{equation}
E_{d}(u)= \frac{k_{d}(u)}{n_{d}(u)}
\end{equation}
\item \emph{Hierarchical common degree} $(H_{d}(u))$ : The average node degree among the nodes in $R_{d}(u)$, considering all edges connected to nodes in the ring.
\begin{equation}
H_{d}(u) = \frac{2e_{d}(u)+k_{d-1}(u)+k_{d}(u)}{n_{d}(u)}
\end{equation}
\end{itemize}

A summary of the hierarchical measurements considered in the present
work are presented in the table bellow.
\\

\begin{tabular}{||p{5.4cm}|c||}
  \hline
  Hier. number of nodes in the \mbox{ring $R_{d}(u)$.} &  $\mathbf{n_{d}(u)}$     \\ \hline
  Hier. number of edges among the nodes in \mbox{the ring $R_{d}(u)$.} & $\mathbf{e_{d}(u)}$ \\ \hline
  Hier. degree of node $(u)$ \mbox{at distance $d$.} & $\mathbf{k_{d}(u)}$   \\ \hline
  Hier. clustering coefficient of node $(u)$ \mbox{at hier. level $d$.} & $\mathbf{cc_{d}(u)}$  \\ \hline
  Convergence ratio of node $(u)$ at \mbox{hier. level $d$.} & $\mathbf{C_{d}(u)}$ \\ \hline
  Intra-ring node degree of node $(u)$ \mbox{at distance $d$.} & $\mathbf{A_{d}(u)}$ \\ \hline
  Inter-ring node degree of node $(u)$ \mbox{at distance $d$.} & $\mathbf{E_{d}(u)}$ \\ \hline
  Hier. common degree of node $(u)$ \mbox{at distance $d$.} & $\mathbf{H_{d}(u)}$    \\ \hline
\end{tabular}
\\

Table 1 -- Summary of the hierarchical measurements considered in this work.

\section{Evaluation of Discriminative Power}
\label{sec:discrim}

Given a complex network, it is possible to organize several selected
measurements of its topology into a \emph{feature vector}
$\overrightarrow{\mu}$ (e.g.~\cite{Costa_Rodrigues:2005}), which
therefore provides a quantitative description of properties of the
network.  Multivariate statistical methods
(e.g.~\cite{CostaCesar:2001}) can then be applied in order to separate
such vectors into clusters or to identify the category of the network.

In a similar fashion, it is possible to assign a feature vector to
individual nodes of the network, so that they can be characterized and
organized into classes.  Although simple measurements such as the node
degree and clustering coefficients can be used for this purpose, they
are generally not enough for a discriminative characterization of
nodes at the individual level because several nodes in a large network
will have identical values of such measurements.  The hierarchical
extensions of the node degree and clustering coefficient, combined
with the ancillary hierarchical measurements described in this work,
account for substantially enhanced discrimination of the local
properties of the connectivity around each node, therefore diminishing
the degeneracy of the description.  In other words, several nodes may
have the same immediate node degree, but it is rather unlikely that
they will also share other hierarchical degrees.  At the same time,
nodes which do present similar connectivity patterns along the
hierarchies can be clustered into meaningful classes by considering
feature vectors composed of hierarchical measurements.

In order to illustrate the above possibilities, we considered a
\emph{S. cerevisiae} protein-protein interaction network $\Gamma$
\cite{Sprinzak:2001} containing $N=1922$ nodes and without
self-connections and isolated nodes.  A perturbed version of this
network was obtained by rewiring the edges with probability $p$.
Nodes in these two networks are then characterized in terms of several
combinations of hierarchical measurements.  In order to quantify the
discriminative power of such measurements, we repeatedly selected a
node from the original network and identified among all nodes of the
perturbed network the node which leads to the smallest Euclidean
distance between the respective feature vectors.  In case these two
nodes are verified to indeed correspond one another (recall that the
identity of the nodes is guaranteed because the perturbed network is
derived from the original network by rewiring), we understand there
has been a correct identification.

\begin{figure*}
 \begin{center}

  \includegraphics[scale=0.17,angle=0]{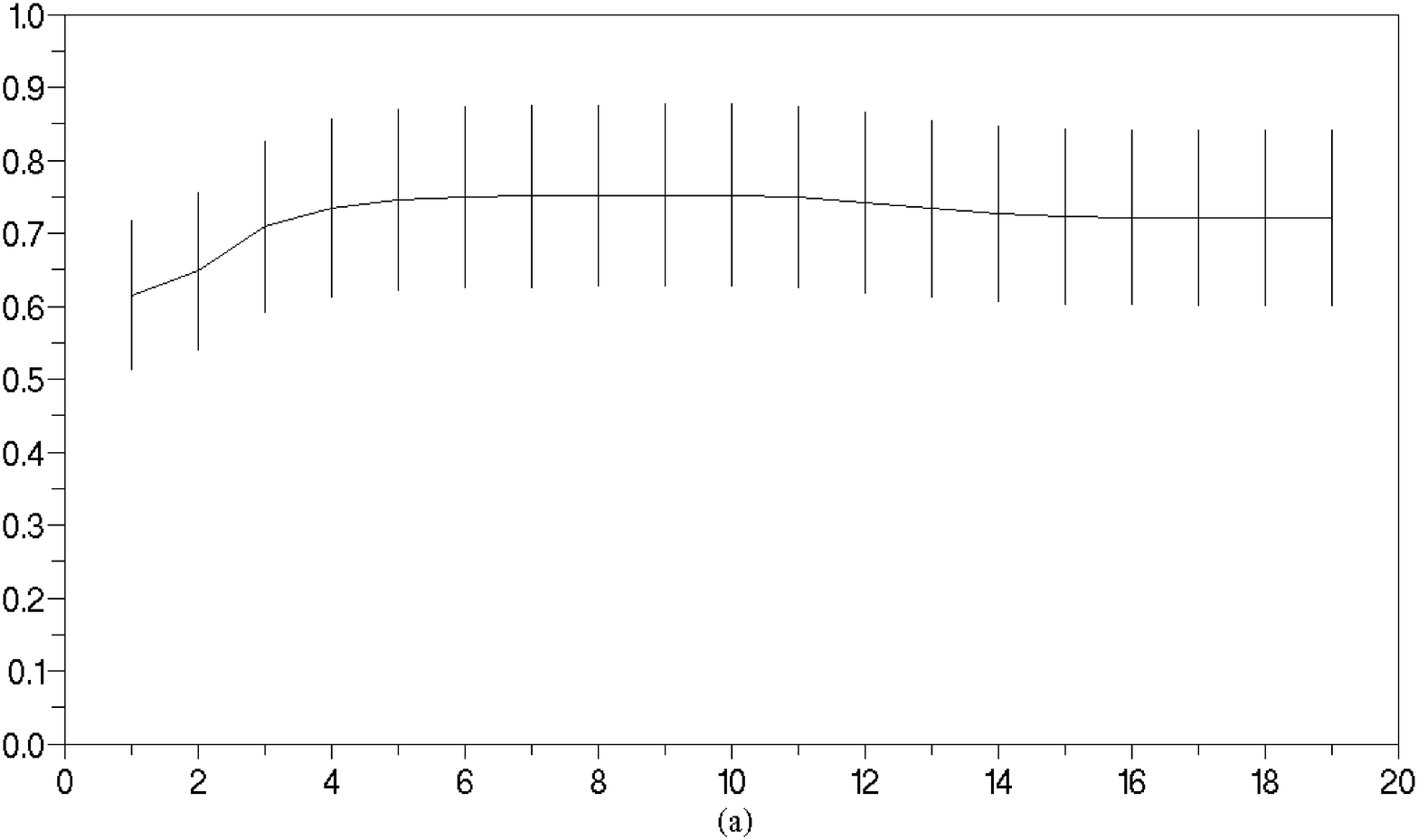}   \hspace{0.2cm}
  \includegraphics[scale=0.17,angle=0]{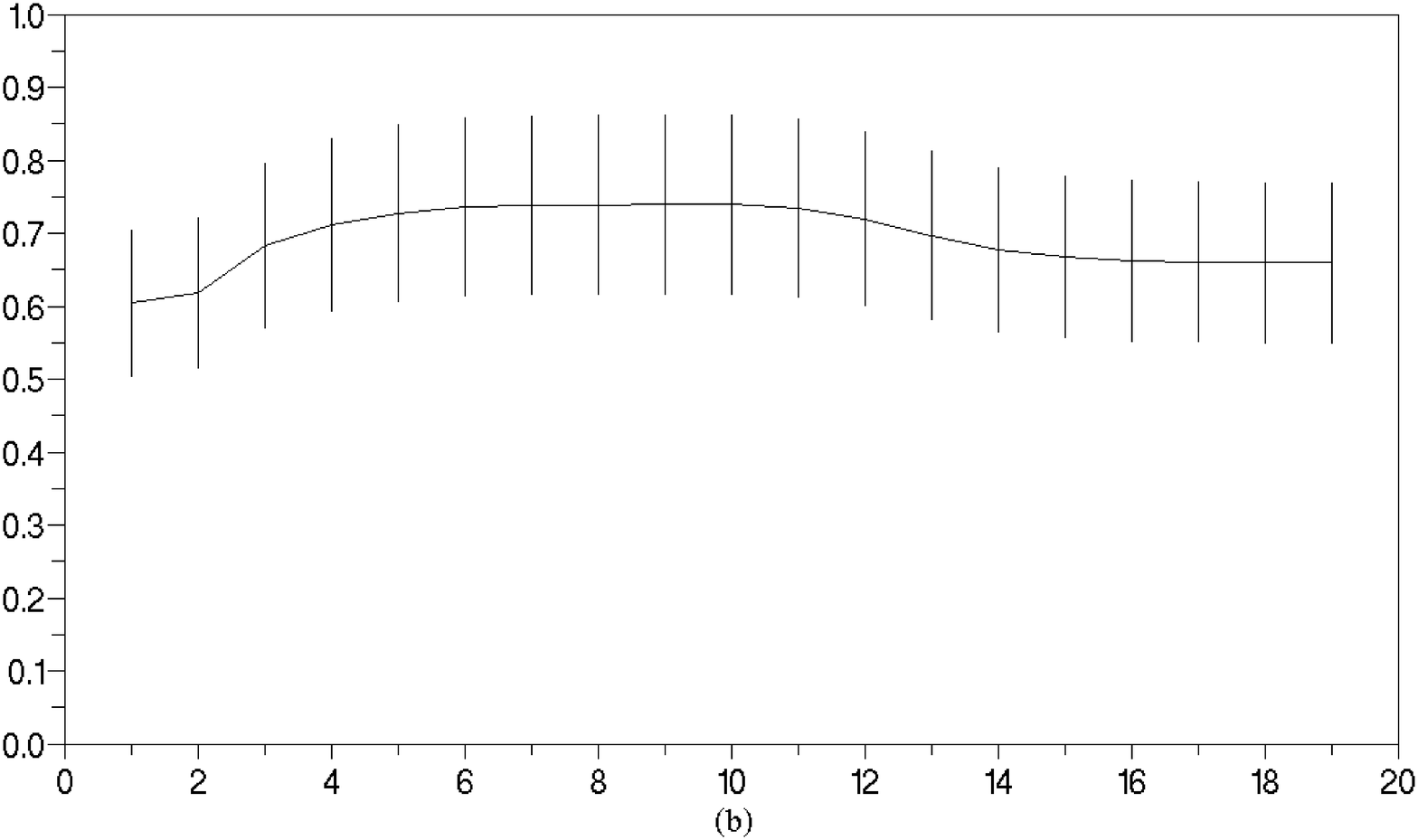}   \hspace{0.2cm}\\
  \includegraphics[scale=0.17,angle=0]{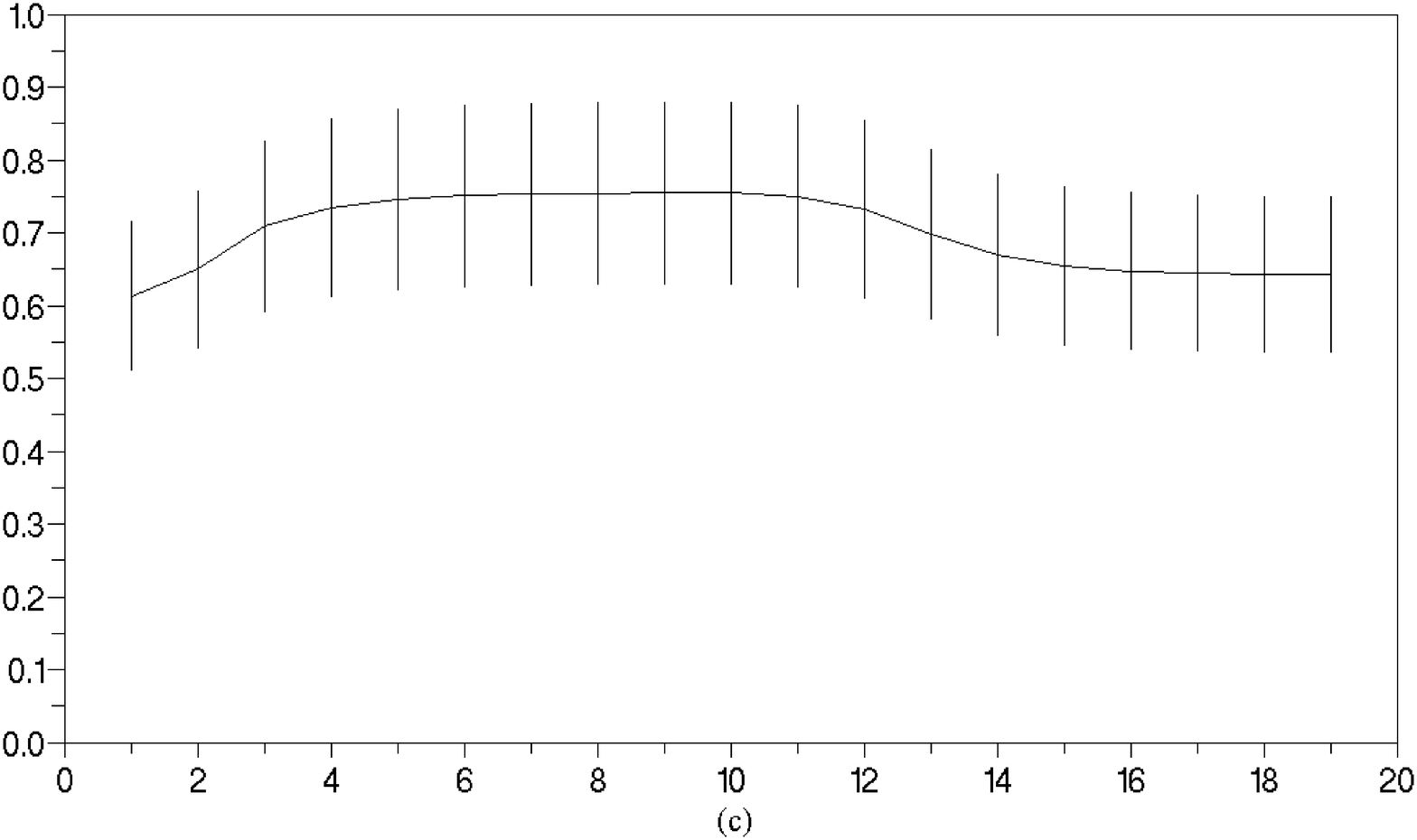}   \hspace{0.2cm}
  \includegraphics[scale=0.17,angle=0]{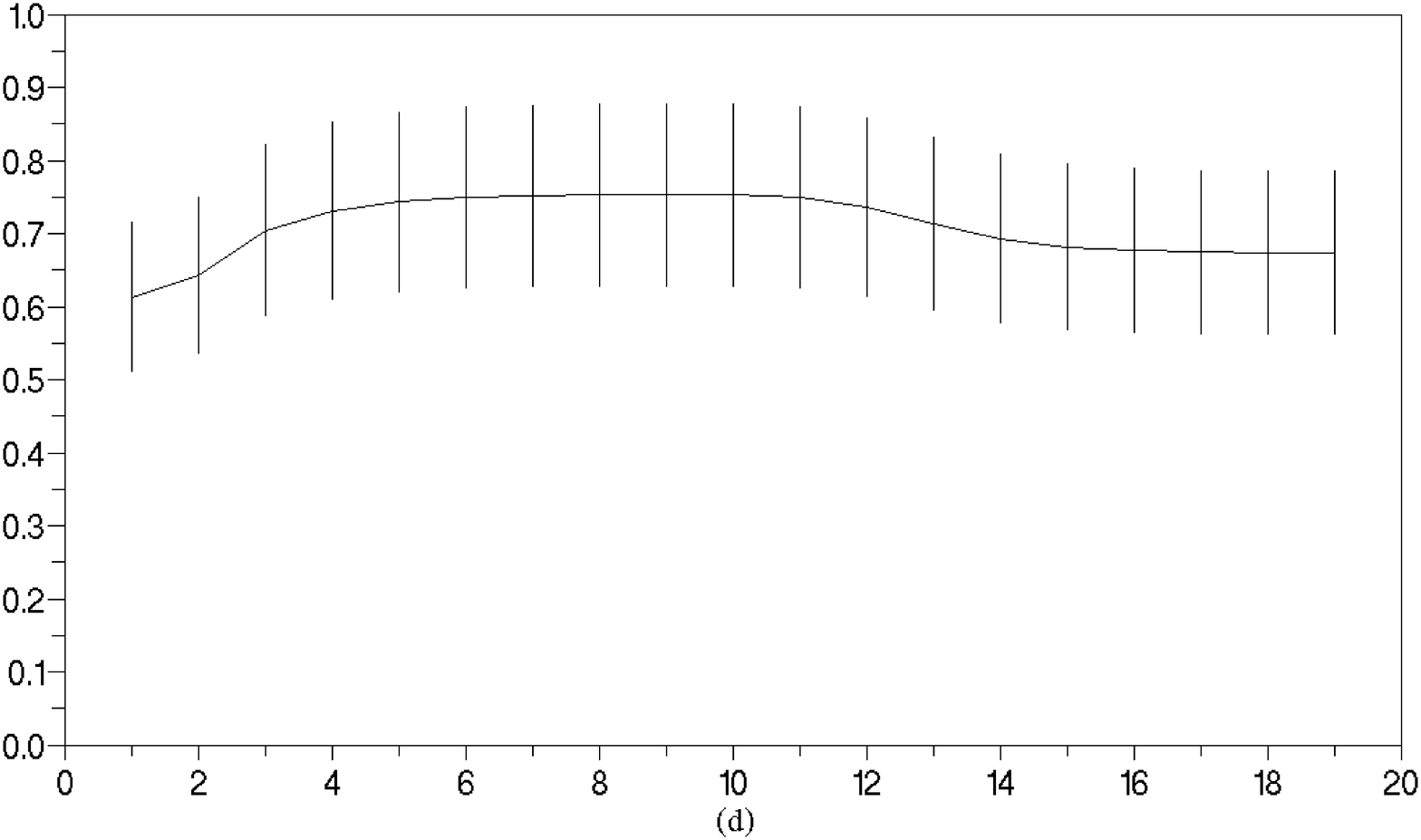}   \hspace{0.2cm}\\
   \caption{ Four hierarchical measurements combinations for the
   feature vector: (a) $A_{d}$ and $E_{d}$; (b) $E_{d}$ and $H_{d}$;
   (c) $C_{d}$, $A_{d}$ and $H_{d}$ and (d) $A_{d}$ and $H_{d}$. \label{fig:det}}

\end{center}
\end{figure*}

Among the several combinations of measurements, considering
varying hierarchical levels, the best results were obtained for
pairwise combinations of $A_{d}$, $E_{d}$, $H_{d}$ and $C_{d}$,
particulalry the four situations shown in Figure~\ref{fig:det}.
The diagrams in this figure depict the average $\pm$ standard
deviation of the percentage of correctly identified nodes by using
the identified pairs of measurements up to the hierarchical levels
identified in the x-axis (varying from 1 to 19).  A number of
interesting features can be identified from such results.  First,
it is interesting to notice that the average of correct
identifications undergoes the three following regimes: (i
increases along the 4 or 5 initial hierarchies; (ii) stays nearly
constant until about 12 hierarchical levels, and (iii) then
decreases steadily.  This behavior is observed for all graphs in
Figure~\ref{fig:det}.  The average performance increase in (i) is
a direct consequence of the fact that more information about the
network connectivity around each node is being taken into account.
The performance plateau and decrease taking place after 4 or 5
hierarchical levels are considered for the measurements reflects a
degeneration in the discriminative power of the measurements
caused by the fact that most of the nodes have been considered at
such hierarchical depths.  Similar performances are observed for
the four cases illustrated in Figure~\ref{fig:det}, with slightly
better results being achieved for the measurement combinations in
(a) and (d). The standard deviations values tend to follow the
mean, with higher variations being observed along the plateaux.

\section{Concluding Remarks}
\label{sec:8}

This article has addressed several issues regarding the generalization
of complex networks measurements.  First, we have shown for the first
time that complex networks and their properties can be formalized in
terms of mathematical morphology, allowing the definition of a series
of measurements such as the generalized versions of the node degree
and clustering coefficient, as well as the possibility to use other
features from mathematical morphology so as to investigate further the
structure of specific subnetworks.  Second, we have emphasized the
importance of identifying and studying the properties of subnetworks
of special interest --- including the set of hubs, outnodes and
3-cycles, and shown that a particularly comprehensive study of such
subnetworks can be obtained by taking into account a whole series of
neighborhoods, as allowed by the the novel proposed concepts of
generalized degrees and clustering coefficient.

While the new set of measurements extended to take into account
subgraphs and hierarchies can be used to derive new network growth
schemes and characterize and classify different types of networks,
they also present power for enhanced discrimination between individual
nodes.  The latter has been illustrated for the first time in this
article with respect to protein-protein interaction networks.  More
specifically, we have shown that the ability to identify
correspondenced between nodes in two versions of a network (in the
case of our example the original and perturbed networks) tend to
increase by considering measurements taking into account multiple
hierarchical levels.  This is a consequence of the fact that the use
of more hierarchical levels allows the measurements to reflect in a
less degenerate way the network connectivity around each node.At the
same time, we have shown that such enhanced discriminative power
tends, with the incorporation of additional hierarchical depths, to
reach a plateau and then to decrease.

The possibilities for future works include the application of the
introduced concepts to community finding, characterization of
resilience to attack, and extensions to measurements aimed at
characterizing the assortative properties of networks.

{\bf Acknowledgments}

Luciano da F. Costa is grateful to FAPESP (process 99/12765-2), CNPq
(308231/03-1), NIH Fogarty, and the Human Frontier Science Program
(HFSP RGP39/2002) for financial support. Luis Enrique Correa da Rocha
is grateful to Human Frontier Science Program for his undergraduate
grant.

\bibliography{Costa}

\end{document}